\documentclass[12pt,twoside,a4paper]{article}
\usepackage[dvips]{epsfig}
\def\gsim{\:\raisebox{-0.5ex}{$\stackrel{\textstyle>}{\sim}$}\:}
\begin{document}
\thispagestyle{empty} 
\title{
\vskip-3cm
{\baselineskip14pt
\centerline{\normalsize DESY 99--125 \hfill ISSN 0418--9833}
\centerline{\normalsize MZ-TH/99--36 \hfill} 
\centerline{\normalsize TTP99-35 \hfill}
\centerline{\normalsize hep--ph/9909230\hfill} 
\centerline{\normalsize August 1999 \hfill}} 
\vskip2cm
Deep Inelastic $e\gamma$ Scattering at High $Q^2$  \\[3ex]
\author{A.~Gehrmann-De Ridder$^{1,2}$, H.~Spiesberger$^3$ and \\[1ex]
P.~M.~Zerwas$^1$ \\[2ex]
{\normalsize $^1$ Deutsches Elektronen-Synchrotron DESY, D-22603
  Hamburg, Germany}\\[0.7ex]
{\normalsize $^2$ Institut f\"ur Theoretische Teilchenphysik,
  Universit\"at Karlsruhe,}\\ 
{\normalsize D-76128 Karlsruhe, Germany} \\[0.7ex]
\normalsize{$^3$ Institut f\"ur Physik,
  Johannes-Gutenberg-Universit\"at,}\\ 
\normalsize{Staudinger Weg 7, D-55099 Mainz, Germany} \\[3ex]
} }
\date{}
\maketitle
\begin{abstract}
\medskip
\noindent
The electromagnetic and weak structure functions of the photon can be
studied in the deep inelastic electron-photon processes $e\gamma \to e
+X$ and $e \gamma \to \nu +X$. While at low energies only virtual photon
exchange is operative in the neutral-current process, additional
$Z$-exchange becomes relevant for high $Q^2$ of order $M_Z^2$ at
$e^+e^-$ linear colliders. Likewise the charged-current process can be
observed at these high energy colliders. By measuring the electroweak
neutral- and charged-current structure functions, the up- and down-type
quark content of the photon can be determined separately.
\end{abstract}

\newpage 

{\bf 1.}  The photon structure functions, which can be measured in
deep-inelastic electron-photon scattering \cite{1}, are an area of
theoretically exciting QCD phenomena. In contrast to the structure
function of the proton, the transverse structure function of the photon
is predicted to rise linearly with the logarithm of the momentum
transfer $Q^2$ and to increase with increasing Bjorken $x$~\cite{2}. The
quark-parton prediction is renormalized in QCD by gluon bremsstrahlung
to order unity without altering the basic characteristics, and the
absolute magnitude of the photon structure function is asymptotically
determined by the QCD coupling~\cite{3}.  The characteristic features of
the photon structure function $F_2^{\gamma}$ are borne out by
experimental data~\cite{3A} though significant improvement of the
experimental accuracy is still demanded.

\begin{figure}[hpb]
\unitlength 1mm
\begin{picture}(120,45)
\put(55,0){\epsfig{file=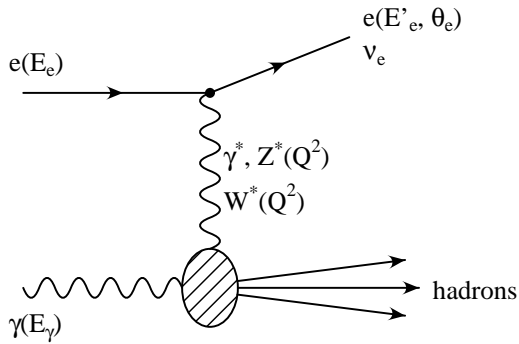,,width=7cm}}
\end{picture}
\caption{\it Kinematics of deep-inelastic electron-$\gamma$ scattering.} 
\label{fig:kine}
\end{figure}

At large momentum transfer $Q^2$ of the order of the $Z$, $W$
masses squared, the neutral-current (NC) $e\gamma$ process becomes
sensitive to the exchange of virtual $Z^*$ bosons in addition to the
virtual $\gamma^*$ photon; moreover the charged-current (CC)
process, mediated by virtual $W^*$ exchange, becomes experimentally
accessible \cite{4,4A}:
\begin{eqnarray}
e + \gamma & \stackrel{\gamma^*, Z^*}{\longrightarrow} & e + X 
\label{eq:1}\\ 
e + \gamma & \stackrel{W^*}{\longrightarrow} & \nu + X 
\label{eq:2}
\end{eqnarray}
These processes can be studied at $e^+e^-$ linear colliders~\cite{5}
where the high energy $\gamma$ beams can be generated by Compton
back-scattering of laser light~\cite{6}.  When exploring the quark
structure of the photon by virtual photons, up- and down-type quarks are
probed at a fixed ratio 4:1. However, since the weak charges of the
quarks differ from the electric charges, the mechanisms (\ref{eq:1}) and
(\ref{eq:2}) can be exploited to determine the up- and down-type quark
content of the photon separately at high $Q^2$.  These measurements are
an interesting task since for asymptotic energies the up/down quark
decomposition of the photon can be predicted by QCD.

{\bf 2.}  The differential cross section of the NC process for
unpolarized leptons and photons is parametrized by two structure
functions $F_2^{\gamma,{\rm NC}}$ and $F_L^{\gamma,{\rm NC}}$:
\begin{equation}
\frac{d^2\sigma}{dxdQ^2} \Bigl[ e^{\pm} + \gamma \rightarrow e^{\pm}+X
\Bigr] = \frac{2\pi \alpha^2}{Q^4} \frac{1}{x} 
\left\{ \left[1+(1-y)^2\right] F_2^{\gamma,{\rm NC}} 
- y^2 F_L^{\gamma,{\rm NC}} \right\}
\label{eq:3}
\end{equation}
while the CC process is parametrized by three structure functions 
$F_2^{\gamma,{\rm CC}}$, $F_L^{\gamma,{\rm CC}}$ and $F_3^{\gamma,{\rm
    CC}}$: 
\begin{eqnarray}
\lefteqn{
\frac{d^2\sigma}{dxdQ^2} \Bigl[e^{\pm} +\gamma  \to 
\stackrel{{\scriptscriptstyle (-)}}{\nu} +X \Bigr] =}
\nonumber\\
& &  \frac{G_F^2}{4\pi} \left[\frac{M_W^2}{Q^2+M_W^2} \right]^2 
\frac{1}{x} \left\{ \left[1+(1-y)^2\right]
F_2^{\gamma,{\rm CC}} \mp \left[1-(1-y)^2\right]   
   x F_3^{\gamma,{\rm CC}} 
- y^2 F_L^{\gamma,{\rm CC}} \right\}
\label{eq:4}
\end{eqnarray}
$\alpha$ is the electromagnetic coupling defined at the scale $Q^2$,
$G_F$ the Fermi coupling; $M_W$ and $\sin^2\theta_w$ are the $W$ mass
and the electroweak mixing angle. The momentum transfer $Q^2$, the
Bjorken variable $x=Q^2/2q\cdot p_\gamma$ and $y=q\cdot p_\gamma/k\cdot
p_\gamma$ (with $p_{\gamma}$, $q$ denoting the $\gamma$, $\gamma^{\ast}$
and $k$ the incoming lepton four-momenta) can be expressed in terms of
the electron energies $E_e$, $E'_e$, the scattering angle of the
electron $\Theta_e$ and the invariant hadron energy $W_h$ in the final
state for the NC process:
\begin{eqnarray}
{\rm \bf NC:}\hspace{1cm}  Q^2 &=& 2 E_e E'_e (1-\cos \Theta_e) \nonumber \\
x &=& Q^2 /(Q^2 +W_h^2) \nonumber \\
y &=& 1-E'_e/E_e \cos^2\Theta_e/2\nonumber 
\end{eqnarray}
For the CC process, including the invisible neutrino in the final state,
the Blondel-Jacquet representation of these variables must be used, defined
in terms of hadron momenta in the final state~\cite{7}:
\begin{eqnarray}
{\rm \bf CC:} \hspace{1cm} Q^2 &=& p_{\perp h}^2/(1-y) \nonumber \\
x &=& Q^2 /(Q^2 +W_h^2) \nonumber \\
y &=& (E_h-p_h)/E_h \hspace{11mm}\nonumber  
\end{eqnarray}
$E_h$, $p_h$ and $p_{\perp h}$ are the sum of hadron energies and of
longitudinal and transverse momenta in the laboratory frame with respect
to the beam axis. The analysis of the structure functions does not
depend on the $\gamma$ spectrum of the initial state in either case; in
fact, the invariant ($e\gamma$) energy can be reconstructed on an
event-by-event basis from the momentum transfer and the scaling
variables, $s_{e\gamma}=Q^2/(xy)$.  Experimental observation of the
events requires minimal scattering angles in the NC process and hadron
angles in the final state of the CC process.  In practice this restricts
$Q^2$ to more than $50$ GeV$^2$.

\begin{figure}[t]
\unitlength 1mm
\begin{picture}(120,75)
\put(25,-5){\epsfig{file=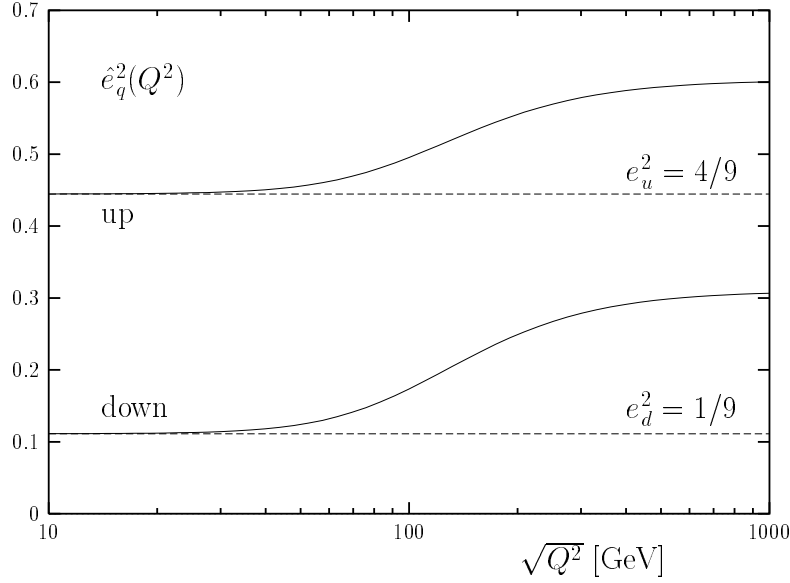,width=11cm}}
\end{picture}
\caption{\it The up and down effective charges as a function of
  $\sqrt{Q^2}$.}  
\label{fig:chargeeff}
\end{figure}

{\bf 3.} Restricting ourselves in this letter\footnote{In an extended
  version of this paper we will present a complete analysis of QCD
  corrections in next-to-leading order as well as an elaborate account
  of CKM and mass effects for heavy quark final states.} to
leading-order QCD predictions which have been proven in Ref.\ \cite{8}
to be quite accurate, the relevant structure functions $F_2^{\gamma,{\rm
    NC}}$, $F_2^{\gamma,{\rm CC}}$, and $xF_3^{\gamma,{\rm CC}}$ for
electron scattering can be expressed by the densities of the light quark
species $u$,\,.\,. $b$:
\begin{equation}
F_2^{\gamma,{\rm NC}} = 2\sum_q \hat{e}_q^2 x \;q(x,Q^2) \label{eq:5}
\end{equation}
and
\begin{eqnarray}
F_2^{\gamma,{\rm CC}} &=& x\left(u+c+d+s\right)
\nonumber\\
x F_3^{\gamma,{\rm CC}} &=& x\left(u+c-d-s\right) \label{eq:6}
\end{eqnarray}
In this representation, use has been made of $\bar q = q$ for the quark
distributions in the photon. The charged-current structure functions for 
positron scattering are obtained from (\ref{eq:6}) by exchanging up-
and down-type quarks. The effective charges $\hat{e}_q$ are
given by the coherent superposition of the electric $\gamma$ and the 
electroweak left- and right-chiral $Z$ charges:
\begin{equation}
\hat{e}_{q}^2=\frac{1}{4} \sum_{i,j=L,R} \left[ e_{q} - 
\frac{Q^2}{Q^2 +M^2_Z} 
\frac{z_i(e)z_j(q)}{\sin^2\theta_w \cos^2\theta_w} \right]^2
\end{equation}
with 
\begin{eqnarray*}
z_L(f) &=& I_{3L}(f)-e_f \sin^2\theta_w \\
z_R(f) &=& ~~~~~~~~ -e_f \sin^2\theta_w 
\end{eqnarray*}
The modification of the electric charge $e_q^2$ to the effective charge
$\hat{e}_q^2$ becomes increasingly important at $Q^2$ above $10^3$
GeV$^2$.  Anticipating that experiments may be feasible up to $Q^2$ of
order of $10^5\;\rm{GeV}^2 $ at 500 GeV colliders (see \cite{5} for
experimental simulations), and of order $10^6\; \rm{GeV}^2$ at 1 TeV
colliders, the $Q^2$ evolution of the effective charges $\hat{e}^2_u$
and $\hat{e}^2_d$ in the transition zone is illustrated in Fig.\ 
\ref{fig:chargeeff}. $\hat{e}^2_u$ evolves from 4/9 asymptotically to
0.60 and $\hat{e}^2_d$ from 1/9 to 0.31.

\begin{figure}[hp]
\unitlength 1mm
\begin{minipage}[t]{16cm}
\begin{picture}(120,80)
\put(25,80){\epsfig{file=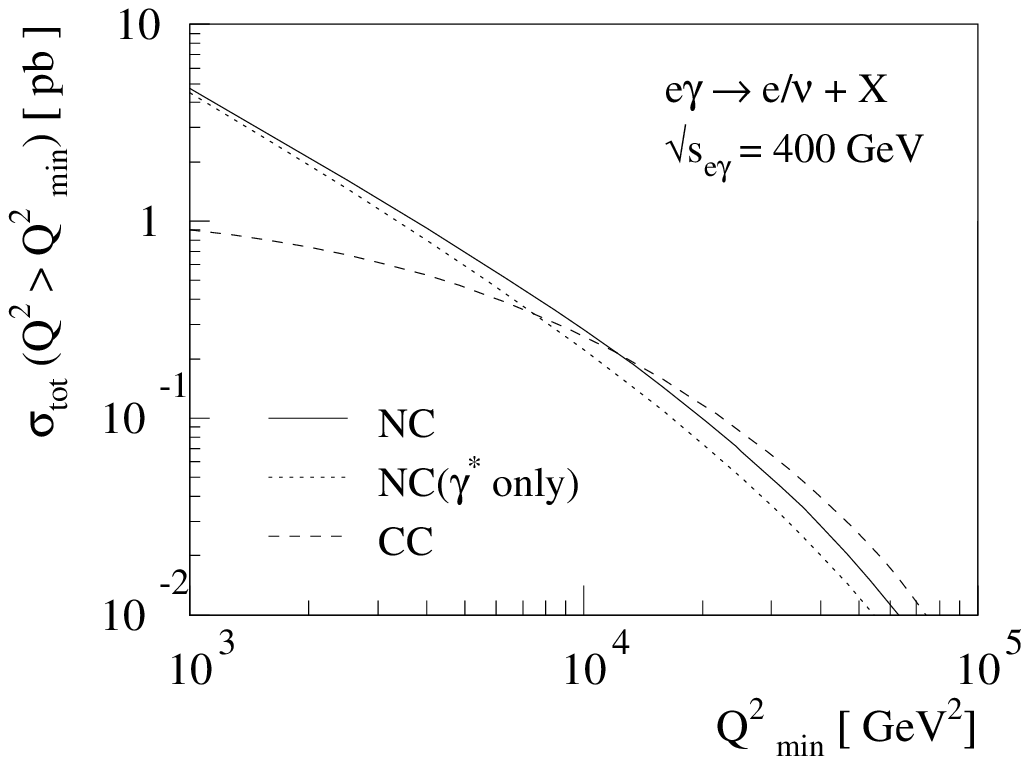,angle=-90,width=11cm}}
\end{picture}
\begin{center}
\label{fig:sig500xg}
\caption{\it The deep inelastic $e\gamma$ cross sections as a function of 
$Q_{\rm {min}}^2$ for the $e\gamma$ collider energy $\sqrt{s_{e\gamma}} 
= 400$ GeV; parton densities: Ref.\ \cite{9}.}
\end{center}
\end{minipage}
\begin{minipage}[t]{16cm}
\begin{picture}(120,80)
\put(25,80){\epsfig{file=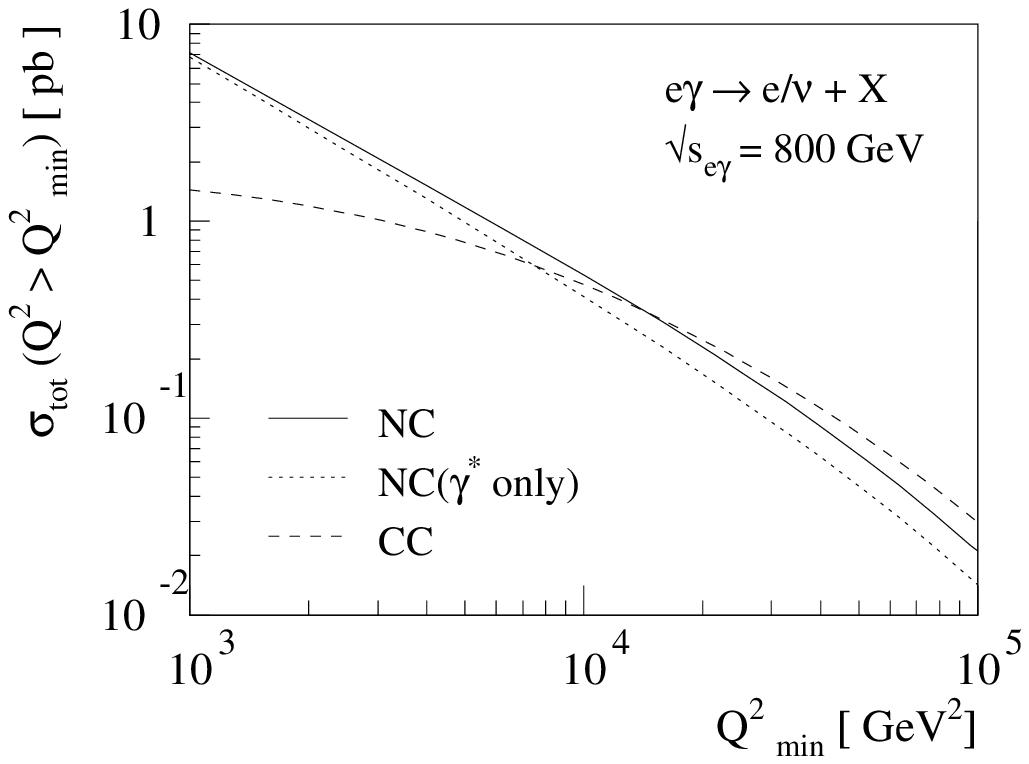,angle=-90,width=11cm}}
\end{picture}
\begin{center}
\label{fig:sig1000xg}
\caption{\it The deep inelastic $e\gamma$ cross sections as a function of 
$Q_{\rm{min}}^2$  for the $e\gamma$ collider energy $\sqrt{s_{e\gamma}} 
= 800$ GeV; parton densities: Ref.\ \cite{9}.}
\end{center}
\end{minipage}
\end{figure}

Based on the lowest-order QCD representation, the NC and CC cross
sections for electron and positron scattering can be written in a simple
form in the variables $x$, $y$ \cite{4}:
\begin{eqnarray}
\frac{{\rm d}\sigma}{{\rm d}x {\rm d}y}[e \gamma \to e X] &=&
\frac{2 \pi \alpha^2 s_{e \gamma}}{Q^4}  \left[1 +(1-y)^2 \right]
\sum_q \hat{e}^2_q \;x\; q(x,Q^2) \\
\frac{{\rm d}\sigma}{{\rm d}x {\rm d}y}[e \gamma \to \nu X] &=&  
\frac{G^2_F s_{e \gamma}}{2\pi} \left[\frac{ M^2_W}{M^2_W + Q^2}\right]^2
x \left[(u+c) + (1-y)^2 (d+s)\right] 
\end{eqnarray}
The cross sections for electron and positron beams are equal by
{$\mathcal{CP}$} invariance.  

The size of the cross sections is shown in Fig.\ 3 and Fig.\ 4 for the
NC and CC processes.  The differential cross sections (8, 9) are
integrated down to a minimum $Q^2_{min}\le xy s_{e \gamma}$ for
fixed\footnote{Choosing the laser and electron/positron beams to have
  opposite helicities $\lambda_e \lambda_{\gamma} = -1/2$
  \cite{6,kuehn}, and separating spatially the conversion from the
  reaction point, the $\gamma$ spectrum in Compton backscattering of
  laser light can be tuned sharply with small band width near
  $E_{\gamma}/E_e \simeq 0.8$ for proper laser colors.  The high-energy
  $\gamma$ beam is unpolarized when the initial laser and lepton
  helicities are flipped at the same time.}  initial $\gamma$ energies
$E_{\gamma}=0.8 E_e$ and electron beam energies $E_e= 250$ GeV and 0.5
TeV \cite{5}.  The full curves correspond to the NC cross section
obtained with the parametrization of Ref.\ \cite{9} in which the parton
densities are evolved from an ansatz defined at low $Q^2$; the dotted
lines indicate the electromagnetic component of the NC cross section.
Due to the $Z$ exchange, the cross section increases by $25\,\%$ for
$Q^2_{\rm min} = 5 \times 10^4$ GeV$^2$ at $\sqrt{s_{e\gamma}} = 400$
GeV. The dashed lines correspond to the CC cross section. Similar
predictions follow from the parametrization \cite{grs}.

\begin{figure}[bhp]
\unitlength 1mm
\begin{picture}(120,80)
\put(25,-5){\epsfig{file=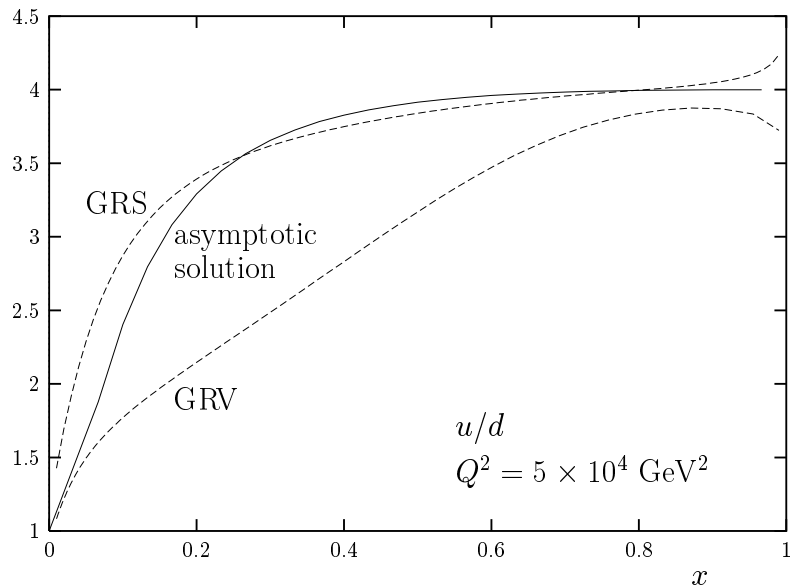,width=11cm}}
\end{picture}
\caption{\it The ratio of the $u$ to $d$ quarks content in the photon as 
  a function of $x$ at $Q^2= 5\times 10^4\; \rm{GeV}^2$ using the
  leading-order parametrizations of Ref.\ \cite{9} (GRV) and Ref.\
  \cite{grs} (GRS). The asymptotic solution \cite{3} is indicated by the
  full line.}
\label{fig:udratio}
\end{figure}

{\bf 4.} Examining the predicted size of the cross sections in Fig.\ 3
and Fig.\ 4, it is obvious that the NC and CC processes can be
investigated for an experimentally large range of $Q^2$. With the
expected $e \gamma$ luminosity in excess of $\int L \gsim 100fb^{-1}/a $
in a high-luminosity machine, such as TESLA \cite{11}, a cross section
of size $\sigma \sim 10^{-1}$ pb gives rise to a sample of $\sim 10^4$
events per year.  The effect of the $Z$ exchange on the NC process and
the CC process can be exploited to separate the up- and down-type quark
components of the photon, as evident from the cross sections (8, 9). The
separation will allow to examine the initial conditions for the
evolution of the parton densities from low $Q^2$ values. Choosing an
incoherent instead of a coherent superposition of vector-meson states
leads to quite different $u/d$ ratios. While in the valence region the
ratio of $u$ to $d$ quarks in the photon is expected to be 4:1, the
ratio should approach 1:1 when the evolution to democratically
distributed sea quarks is fully operative. Involving a two-step
splitting process $q \to g\to q'$, the transition can only be observed
for sufficiently small $x$ as exemplified in Fig.\ \ref{fig:udratio} for
the parametrizations of Refs.\ \cite{9,grs} at $Q^2$ = $5\times 10^4$
GeV$^2$ and the asymptotic solution. $x$ values down to order $10^{-2}$
will be accessible at these colliders even for large $Q^2$ values.

\subsection*{Acknowledgements}
We thank D.\ J.\ Miller for very useful discussions on experimental
aspects of this analysis, and we thank G.\ Blair for valuable comments
on the manuscript. A.\ G.\ is grateful to A.\ Wagner for financial
support during her stay at DESY where part of this work was carried out.


\end{document}